\documentclass{article}

\PassOptionsToPackage{numbers, compress}{natbib}
\usepackage{enumitem}

\usepackage[preprint]{neurips_2025}

\usepackage[utf8]{inputenc} 
\usepackage[T1]{fontenc}    
\usepackage{hyperref}       
\usepackage{url}            
\usepackage{booktabs}       
\usepackage{amsfonts}       
\usepackage{nicefrac}       
\usepackage{microtype}      
\usepackage{xcolor}         
\usepackage{graphicx}
\usepackage{subcaption}     
\usepackage{amsmath}
\usepackage{caption}         

\title{Helix Parallelism: Rethinking Sharding Strategies for Interactive Multi-Million-Token LLM Decoding}

\author{%
  Nidhi Bhatia
  \And
  Ankit More \\
  \And
  Ritika Borkar \\
  \And
  Tiyasa Mitra \\
  \And
  Ramon Matas \\
  \And
  Ritchie Zhao \\
  \And
  Max Golub \\
  \And
  Dheevatsa Mudigere \\
  \And
  Brian Pharris \\
  \And
  Bita Rouhani \\
  \AND
  \textnormal{NVIDIA Corporation}
}

\begin{document}

\maketitle

\begin{abstract}
  As LLMs scale to multi-million-token KV histories, real-time autoregressive decoding under tight Token-to-Token Latency (TTL) constraints faces growing pressure.
  Two core bottlenecks dominate: accessing Feed-Forward Network (FFN) weights and reading long KV caches.
  While Tensor Parallelism (TP) helps mitigate the cost of FFN weight reads, it does not scale well for attention.
  When TP width exceeds the number of KV heads, it leads to inefficient KV duplication, limits parallelism, and constrains batch size.
  Simultaneously, DRAM reads for long KV histories scale linearly with batch size, further capping efficiency.
  
  We introduce \textbf{Helix Parallelism}, a hybrid execution strategy that applies KV parallelism during attention to shard KV caches across GPUs, then reuses the same GPUs for TP in dense LLMs or TP$\times$Expert Parallel (EP) in MoEs during FFN computation.
  To preserve exact attention behavior, Helix includes a lightweight communication step.
  To minimize the exposed communication cost, we introduce \textbf{Helix HOP-B}.
  Helix HOP-B effectively minimizes communication overhead through batchwise overlap, preserving low TTL while improving GPU efficiency.
  Compared to conventional parallelism approaches, Helix reduces TTL by up to 1.5x at fixed batch sizes and supports up to 32$\times$ larger batches under the same latency budget for DeepSeek-R1, pushing forward the throughput-latency Pareto on Blackwell and making real-time inference with ultra-long-sequence practical.
\end{abstract}

\section{Introduction}
Large Language Models (LLMs) are increasingly expected to handle ultra-long histories~\cite{gemini15}~\cite{llama4}, precomputed context spanning millions of tokens\footnote{Here, context refers to the sequence of previously generated tokens, whose intermediate key and value representations are stored as KV cache and accessed at every decoding step.}, while still delivering millisecond-level Token-to-Token Latency (TTL) for interactive applications. Long contexts allow models to maintain narrative coherence, capture long-range dependencies, and support complex reasoning and planning. At the same time, interactivity demands that each token be generated almost instantly, as users expect real-time responsiveness from AI assistants, copilots, and autonomous agents.

Decoding under these dual pressures exposes two fundamental bottlenecks.
First, \textbf{KV cache reads during self-attention} become increasingly expensive.
The KV cache size grows linearly with both context length and batch size, rapidly overwhelming DRAM capacity and bandwidth.
To fit multi-million-token caches, systems are often forced to reduce batch sizes, but read time still remains high with longer histories, driving TTL beyond acceptable limits.
Second, \textbf{FFN weight reads during autoregressive decoding} contribute heavily to latency.
Generating every new token requires loading large feed-forward network (FFN) weights from DRAM.
With small batch sizes, this cost cannot be amortized, making weight reads another dominant factor in overall decoding time.

Modern attention variants such as Grouped-Query Attention (GQA)~\cite{ainslie2023gqatraininggeneralizedmultiquery}, Multi-Query Attention (MQA)~\cite{shazeer2019mqa}, 
and Multi-Head Latent Attention (MLA)~\cite{deepseekai2024deepseekv2strongeconomicalefficient} reduce KV-cache pressure by merging multiple keys and values into shared representations, i.e., collapsing $Q$ original query heads into $K$ KV heads where $K<Q$. A typical value of $K$ is $8$ or less in modern LLMs. For MLA, there is just a single latent representation of both K and V for all Q heads.

Tensor Parallelism (TP)~\cite{shoeybi2020megatronlmtrainingmultibillionparameter} shards both FFN weights and attention heads evenly over $TP$ GPUs, so that each device only holds and reads its portion of the weights and KV cache. When $TP \le K$, TP naturally splits the KV cache without duplication.
However, once $TP > K$ (common in large models needing high parallelism) each additional shard must store a full copy of the KV cache to serve its assigned query heads, 
despite splitting computation. Beyond this point, increasing $TP$ neither shrinks per GPU KV cache size nor speeds up KV cache reads, imposing a hard ceiling on attention time (see Figure~\ref{fig:combined}) and forcing smaller batch sizes under tight TTLs.
Furthermore, since TP must be capped at $K$ shards (the number of KV heads), only those $K$ GPUs can be used to shard the FFN, which accounts for roughly two‐thirds of 
model parameters. Reading these large weight matrices on just $K$ devices not only monopolizes memory that could host additional KV caches, 
but also makes FFN weight loads the primary latency bottleneck, further limiting how many concurrent batches can be maintained in real‐time decoding.

To tackle KV cache scaling, recent work like Medha~\cite{agrawal2025medhaefficientlyservingmultimillion} shards the KV cache across an auto-scaled pool of $N$ GPUs using 
KV Parallelism (KVP), so each device stores only a fraction of the multi-million-token history. 
This approach significantly reduces both per GPU cache size and read latency during self-attention. 
However, Medha and similar methods then gather the attention outputs onto a fixed group of TP GPUs (e.g., 8) for all subsequent FFN computations. 
In effect, while KVP fans out computation across $N$ GPUs for attention, it does not repurpose those same GPUs to further accelerate FFN execution. 
As a result, FFN weight loads remain a latency bottleneck, and hardware resources become increasingly underutilized as $N$ grows.

This paper introduces \textbf{Helix Parallelism} and \textbf{Helix HOP-B} for interactive multi-million-token LLM decoding. Helix parallelism is a hybrid sharding strategy where 
we disaggregate the mapping of attention and FFNs in a temporal pipeline to address both KV cache and FFN bottlenecks. Specifically:
\begin{enumerate}[leftmargin=*]
  \item  \textbf{Attention phase:} Applies KVP to shard the KV cache along the sequence dimension across KVP GPUs, combined with TP across KV heads when $TP \le K$, resulting in a total of $N = KVP\times TP$ GPUs handling attention with no cache duplication.
  \item \textbf{FFN phase:} Immediately reconfigures the same $N$ GPUs, now running either dense TP or combined TP\,$\times$\,Expert Parallelism (in MoE models), to shard FFN weight matrices and accelerate weight reads. This reconfiguration also allows TP widths for FFN to exceed the number of KV heads without reintroducing cache duplication during the attention phase.
\end{enumerate}

HOP-B is an overlap optimization that mitigates communication latencies introduced by Helix Parallelism by masking them behind computation, effectively maintaining low TTL.

By aligning the parallelism scheme to each stage's distinct computational demands, Helix reduces TTL by up to 50\% at fixed batch sizes, and expands batch size by up to 32$\times$ under the same latency budget. Such a hybrid optimization is particularly important on modern GPU systems like Blackwell~\cite{blackwell2024}, which feature large NVLink domains.
Helix parallelism is fully compatible with modern LLM architectures, including MLA and GQA attention, as well as MoEs.

\section{Helix parallelism}
\begin{figure}[htbp]
    \centering
    \includegraphics[width=\textwidth]{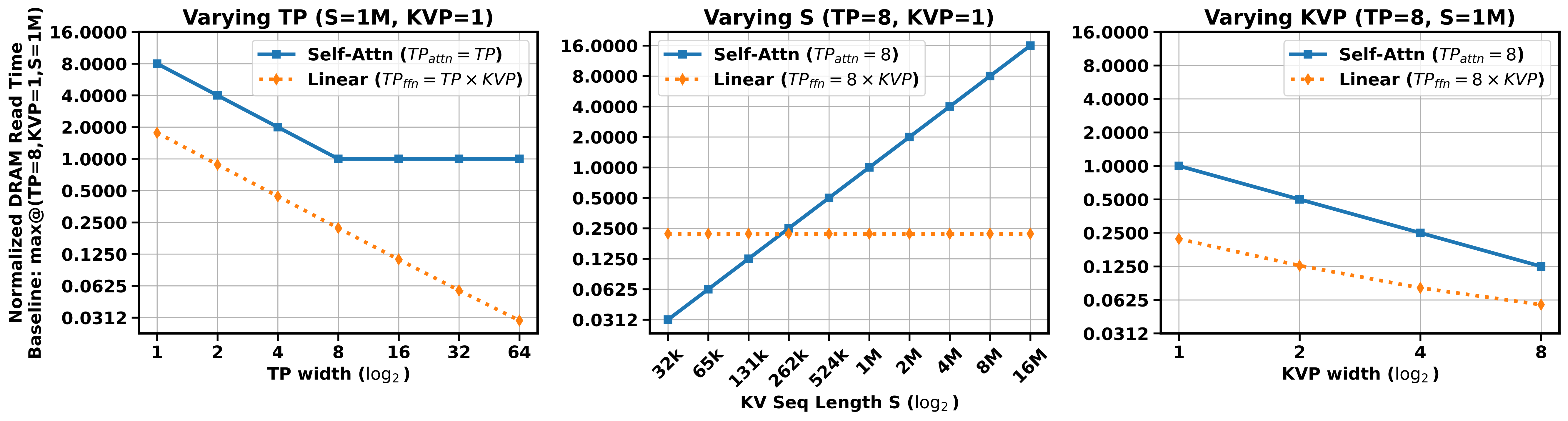}
    \caption{%
    \textbf{Roofline analysis for KV cache and Linear weight reads}, assuming a Dense LLM with batch $B{=}8$, Query heads $Q{=}128$, KV heads $K{=}8$, head size $Hsz{=}128$, and FFN dimension $F{=}65536$ running 
    on GB200 NVL72. Both weights and KV cache, are stored and fetched in FP4. Communication overhead from TP and KVP is not included; these plots show only the change in GPU DRAM‐read latency as TP width and KVP width vary. Details can be found in Appendix \ref{appendixA}
    \textbf{(Left)} DRAM read latency vs. TP width. Benefits plateau beyond $TP{=}K$ due to full KV duplication, highlighting the need for KV sharding in Helix.
    \textbf{(Middle)} DRAM read time vs. KV length $S$. Self-attention cost scales linearly with $S$, eventually dominating latency.
    \textbf{(Right)} DRAM read time vs. KVP width. Helix applies KVP in attention to reduce per-GPU memory traffic and achieve sublinear scaling, enabling multi-million-token inference. The same GPUs are then re-provisioned for TP or EP in FFNs to minimize latency.
    }
    \label{fig:combined}
  \end{figure}

Helix Parallelism is based on a key insight: achieving real-time decoding with multi-million-token KV histories requires decoupling the mapping of attention and FFNs. 
Figure~\ref{fig:combined} provides a high-level intuition for why this separation is essential. 
Helix introduces a temporal pipeline within each layer, allowing the same set of GPUs to be reused across attention and FFN computation, while applying different parallelism strategies 
for each. In the remainder of this section, we describe the core building blocks of Helix Parallelism in detail.

\subsection{Attention partitioning}
\subsubsection{KV partitioning (KVP)}
\begin{figure}[ht!]
    \centering
    \includegraphics[width=\textwidth]{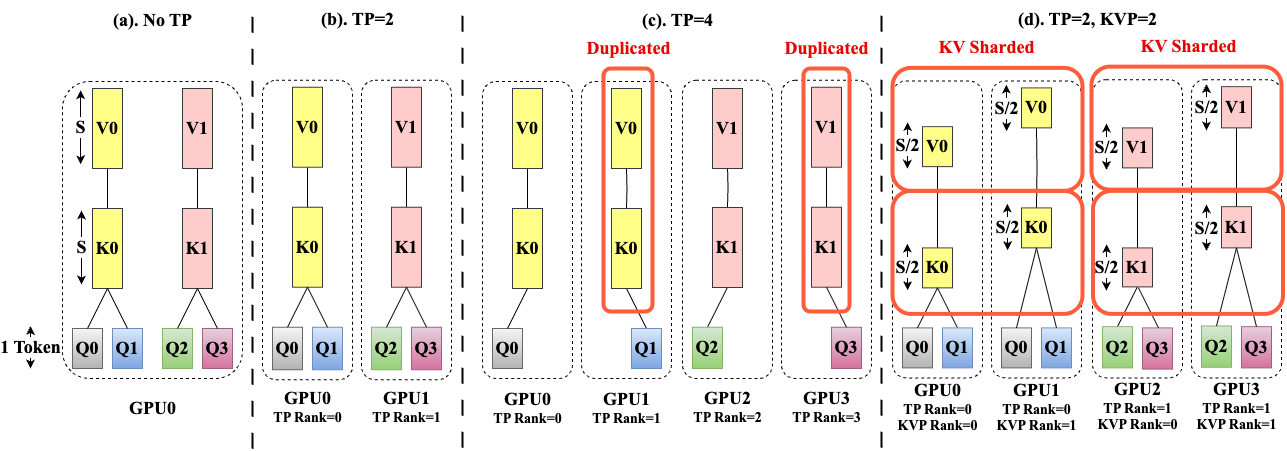}
     \caption{\textbf{Overview of different attention sharding strategies.} 
     Here we are using GQA as an example with Query heads $Q{=}4$ and KV heads $K{=}2$. 
     Each block processes one token with context length S.
     \textbf{(Left)} No TP: all Q and KV heads are co-located on a single GPU; no duplication. 
     \textbf{(Middle-Left)} TP=2: Query heads are split across 2 GPUs; KV heads are still partitioned cleanly since $TP \leq K$. \textbf{(Middle-Right)} TP=4: More shards than KV heads; GPUs must duplicate KV cache to serve their assigned query heads, reintroducing DRAM capacity and bandwidth inefficiencies.
     \textbf{(Right)} Helix (TP=2, KVP=2): Helix shards the KV cache across sequence length ($S$) using KVP, so each KVP rank holds only a slice ($S/2$). By capping TP at $K$ and assigning the remaining GPUs to KVP, Helix avoids duplication and forms a 2D layout: TP splits heads, KVP splits the sequence.
    }
    \label{fig:kv_tp_diagram}
  \end{figure}

  During the attention phase, Helix configures all available GPUs into a pool of $N = \mathrm{KVP}\times\mathrm{TPA}\quad(\mathrm{TPA}\le K)$, then shards the KV cache along the sequence dimension across the $KVP$ GPUs, eliminating full-cache replication and cutting DRAM footprint and bandwidth demands.
  To avoid an expensive pre-attention All-Gather of queries across the KVP GPUs, Helix has each KVP GPU independently compute the full QKV projections.
  Concretely, every GPU takes the full input batch $[B,H]$ and multiplies it by the QKV weight matrices $W_Q \in \mathbb{R}^{H \times (H/TPA)}$, $W_K \in \mathbb{R}^{H \times (\lceil K/TPA \rceil \cdot H_{sz})}$, and $W_V \in \mathbb{R}^{H \times (\lceil K/TPA \rceil \cdot H_{sz})}$ to produce the full QKV projections.
  This means each of the $KVP$ GPUs holds its own full set of query heads, and the corresponding key/value projections, so it can run FlashAttention~\cite{shah2024flashattention3fastaccurateattention} on its KV shard in isolation.
  Each GPU emits a partial attention output and a log-sum-exp scalar per token; a single All-to-All over the query-head axis then exchanges these fragments, and each GPU rescales and sums them to reconstruct the exact softmax-normalized attention in one communication round, with no extra synchronization or normalization passes~\cite{dao2023flashdecoding}.  
  This All-to-All exchange also realigns the data partitions for subsequent TP-based execution. 
  Figure~\ref{fig:kv_tp_diagram} provides a high-level overview of different sharding schemes in attention and their corresponding layout.

\subsubsection{Optimized all-to-all communication}
The communication volume in Helix parallelism is independent of the KV-sequence length $S$ and scales only with the number of query tokens in a batch $B$ and the hidden dimension $H$. 
This constant per-token overhead makes Helix parallelism highly scalable, allowing efficient decode-time attention even with multi-million-token KV caches.

\subsubsection{Batch-wise communication–computation overlap}
\begin{figure}[ht!]
  \centering
  \includegraphics[width=\linewidth]{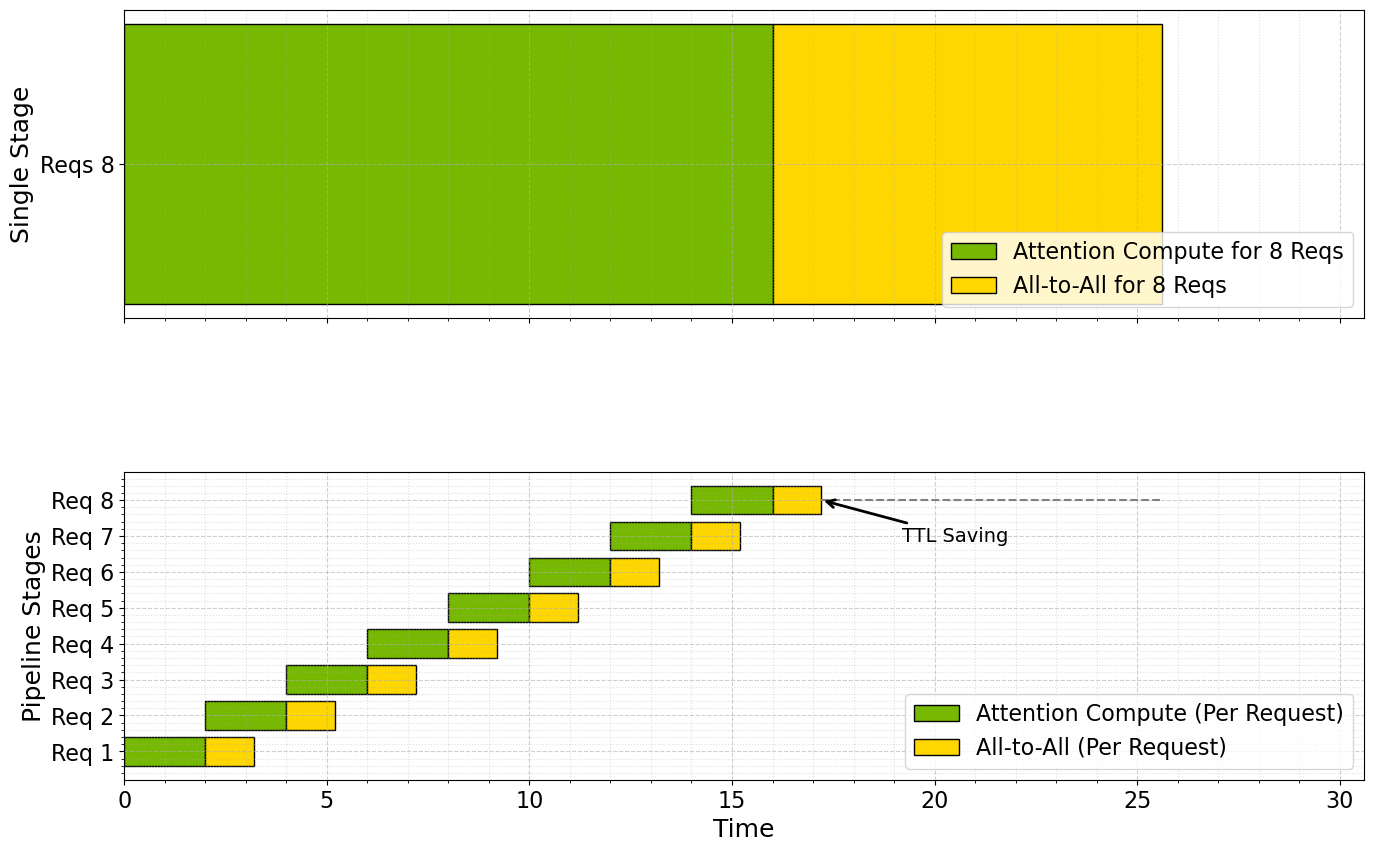}
  \caption{\textbf{KVP All-to-All exposed time:} \textbf{(Top without HOP-B)} All 8 requests execute in lockstep, each consumes 16 time units of attention before initiating 9.6 units of communication, 
  for a total span of 25.6 units with no overlap; \textbf{(Bottom with HOP-B)} Requests are pipelined so that while one request's communication is ongoing, the next request begins its attention compute immediately. Here each request's compute (2 units) and communication (1.2 units) are drawn to scale as discrete blocks. A dashed gray line and "TTL Saving" arrow highlight the reduction in overall TTL from 25.6 units in the baseline down to 17 units when HOP-B is enabled.}
  \label{fig:pipeline}
\end{figure}

To minimize exposed communication time, we introduce and deploy \textbf{Helix HOP-B} (Helix Overlap Pipeline – Batch-wise), a fine-grained pipelining strategy that overlaps All-to-All communication 
with ongoing attention computation across the batch dimension. As illustrated in Figure~\ref{fig:pipeline}, once the attention output for the first query token is computed, 
Helix immediately initiates All-to-All for that token while concurrently processing attention for the next. This overlap maintains high hardware utilization and effectively hides 
communication latency, further reducing TTL for real-time inference decoding.

\subsection[Seamless re-provisioning to TP (dense FFNs) and TP $\times$ EP (MoE FFNs)]{FFN partitioning}
\begin{figure}[ht!]
  \centering
  \begin{minipage}{\linewidth}
    \makeatletter
    \renewcommand{\thempfootnote}{\arabic{mpfootnote}}
    \makeatother

    \includegraphics[width=\linewidth,trim={0 7.5cm 0 7.1cm},clip]{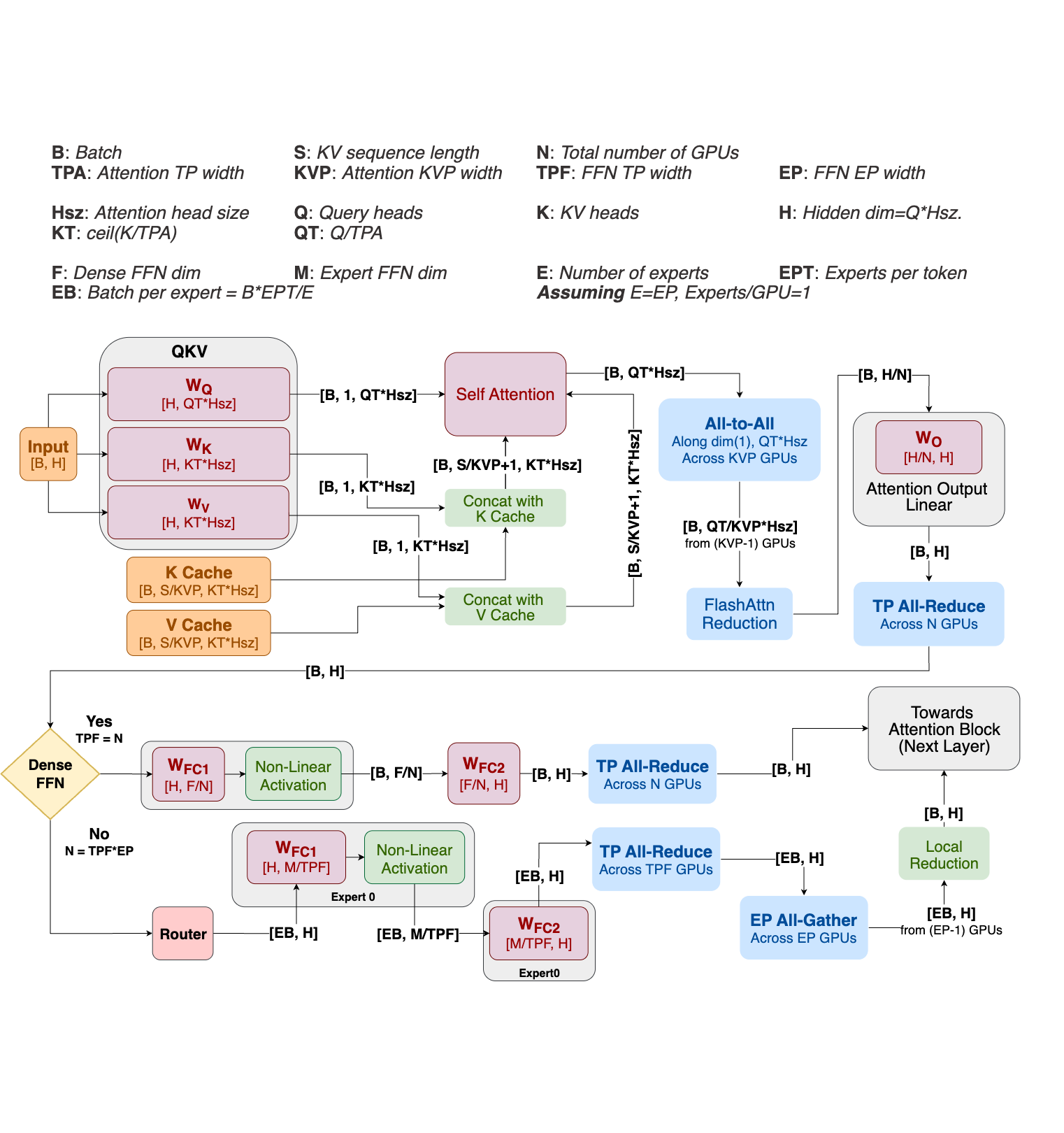}

    \captionof{figure}{\textbf{Helix's layout per-GPU workflow in attention and FFN stages.} Helix reuses the same pool of $N$ GPUs by configuring $N = {KVP}\times{TPA}\quad({TPA}\le K)\;\longrightarrow\;N = {TPF}\times{EP}$ on a per-layer basis. \textbf{(Top)} During attention, each of the $KVP$ GPUs independently projects the full batch $[B,H]$ into QKV, runs FlashAttention~\cite{shah2024flashattention3fastaccurateattention} on its KV shard to produce partial outputs and log-sum-exp scalars, and then participates in a single All-to-All over the query-head axis; each GPU rescales and sums the received fragments into the exact softmax-normalized tensor, followed by a TP All-Reduce to form the final $[B,H]$ attention output. \textbf{(Bottom)} For FFNs, Helix follows two modes depending on model type: for \emph{Dense} FFNs $EP=1$, it retains all $N$ GPUs in tensor-parallel $TPF=N$ to compute $[B,H]\;\to\;[B,F/N]\;\to\;[B,H]$ followed by a TP All-Reduce; for \emph{MoE} FFNs $EP>1$, it repartitions the $N$ GPUs into a $TPF\times{EP}$ grid, routes tokens to the appropriate experts, applies TP to FC layers within each expert group, performs an intra-expert All-Reduce followed by an inter-expert All-Gather, and concludes with a local reduction to yield $[B,H]$. Helix switches between these configurations seamlessly, enabling zero-downtime pipelining and better GPU utilization.
  }
  \label{fig:helix_overview}
\end{minipage}
\end{figure}

After self-attention, Helix reuses the same pool of $N = \text{KVP} \times \text{TPA}$ GPUs, originally configured for KV partitioning and, where applicable, TP, to execute the FFN in a layout optimized for the model type, whether dense or MoE.
Figure~\ref{fig:helix_overview}, provide a detailed end-to-end view of Helix parallelism for a single transformer layer. For simplicity, layer‐normalization and residual‐addition are omitted from the diagram.\footnote{While we depict a two‐stage FFN, \emph{FC1} ($H\times F$) and \emph{FC2} ($F\times H$), modern designs often employ gated variants (e.g.\ an additional $H\times F$ gating matrix) or fuse these into combined projections (e.g.\ $H\times2F$ and $2F\times H$).}

At the end of local attention, each KVP GPU has computed partial attention outputs over its KV shard and respective query heads and the entire batch $B$. 
These are exchanged via a single \texttt{All-to-All} along the query-head dimension: each GPU sends its $B \times \bigl(\tfrac{H}{KVP \times TPA}\bigr)$ partial results to every other GPU in KVP domain and receives the corresponding slices in return. 
After rescaling and summing with the per-shard log-sum-exp statistics, each GPU holds the fully normalized attention outputs for its assigned $\tfrac{H}{KVP \times TPA}$ hidden-dimension slice, but for the entire batch, effectively forming a TP group of 
size $TPA\times KVP = N$.
These normalized outputs feed into the post-attention linear projection, which runs in tensor-parallel fashion across the same $N$ GPUs. Each TP rank holds a shard of the projection weight matrix of shape $\frac{H}{N}\times H$, computes its local matrix-multiply on the full batch inputs of shape $B\times\frac{H}{N}$, and then participates in an \texttt{All-Reduce} over the $N$ GPUs. This All-Reduce aggregates the $B\times H$ partial projections into the full $B\times H$ output, which is then passed through layer normalization and into the FFN block.

The post-attention re-provisioning, from KVP to TP for the post-attention linear projection, is identical in both dense and MoE models. After the post-attention linear projection and \texttt{All-Reduce} over the $N$ GPUs on its output, 
the full pool of $N$ GPUs is re-provisioned for the FFNs (see Figure~\ref{fig:helix_overview}): 
\begin{itemize}[leftmargin=*]
    \item {\textbf{Dense FFNs ($EP = 1$)}}: Keep $TPF=N$, all GPUs collaborate on amortizing weight‐read costs and maintaining low latency on small batches. 
    \item {\textbf{MoE FFNs, ($EP > 1$)}}: Repartition into an optimal $TPF\times EP$ grid, choosing $TPF$ versus $EP$ to best match expert size and quantity.
\end{itemize}

By chaining KVP $\times$ TPA for attention, a TP = $N$ post-attention linear projection, and a flexible TPF $\times$ EP FFN layout, Helix ensures all $N$ GPUs remain fully utilized in a zero-downtime pipeline, maximizing scalability and throughput across attention, projection, and FFN stages.  

\subsection{Distributed KV concatenation strategy}
During decoding, each newly generated token is broadcast to all KVP GPUs so that every device has access to the current query. However, Helix staggers the KV cache updates across KVP ranks to maintain balanced memory growth and ensure each GPU appends KV entries uniformly. Specifically, the system appends KV pairs for a fixed number of decode steps (e.g., 16 tokens) to the shard on KVP Rank 0, then switches to KVP Rank 1 for the next 16 tokens, and so on, cycling through all $N$ KVP ranks in round-robin fashion. This staged KV concatenation guarantees that all KVP GPUs contribute to KV storage regardless of batch size or sequence length, avoiding hot spots and distributing KV cache growth evenly across the pool.

\section{Evaluation}
In this section, we evaluate Helix Parallelism against the best known prior LLM sharding methods on NVIDIA's latest GB200 NVL72 hardware~\cite{blackwell2024} with FP4 precision~\cite{microscaling2023}. 
Rather than focusing on isolated configurations, we characterize the full Pareto frontier between \textbf{throughput per gpu} (tokens/sec/gpu) vs. 
\textbf{interactivity} (tokens/sec/user). Here, user interactivity is measured as reciprocal of decoding TTL (token-to-token latency), representing the rate at which 
new tokens are generated for a single user. 
Throughput per GPU is quantified as the total number of tokens generated per second per GPU, reflecting system-wide efficiency.

To provide a more comprehensive view, we also introduce \textbf{batch scalability}, the maximum number of concurrent user requests that can be sustained under a fixed TTL budget. This metric reflects a system's ability to maintain real-time responsiveness at scale.

\subsection{Experimental setup}
To evaluate the performance of different sharding strategies independent of potential feature gaps in a given software framework, we opt to leverage an in-house high-fidelity 
simulator modeling the latest GB200 hardware. The simulator accounts for both compute and communication costs, including latency from inter-GPU NVLink transfers, 
DRAM bandwidth constraints, and FLOP throughput. All performance numbers are normalized to that of the baseline to focus on the trends as opposed to specific performance claims.

We evaluate Helix Parallelism on two large-scale LLMs representative of dense and MoE architectures:
\textbf{(i) Llama-405B} ~\cite{grattafiori2024llama3herdmodels}, a dense 405B parameter model with 128 query heads and 8 grouped KV heads (i.e., GQA attention).
\textbf{(ii) DeepSeek-R1} ~\cite{deepseekai2025deepseekv3technicalreport}, a 671B parameter MoE model with MLA attention. In MLA, key and value projections are absorbed into a latent space during decoding and are not explicitly materialized, effectively resulting in a single KV head shared across all 128 query heads.
Although these models do not yet natively support million-token contexts, we simulate decode-time inference with KV-cache sequence lengths of \textbf{one million tokens and beyond}. This allows us to analyze system-level bottlenecks and assess the potential of Helix for real-time applications at this scale.

Our baseline search space covers the best‐known partitioning strategies, tensor parallelism (TP), pipeline parallelism (PP), expert parallelism (EP), and vanilla KV partitioning (KVP), alongside a full sweep over batch sizes. Here, EP denotes data‐parallel attention coupled with expert‐parallel FFNs, as adopted in production DeepSeek‐R1~\cite{deepseekai2025deepseekv3technicalreport}. Vanilla KVP refers to the original Medha‐style sharding approach, first introduced in~\cite{agrawal2025medhaefficientlyservingmultimillion}.

Each point on the Pareto frontier corresponds to a unique combination of model partitioning and batch size. For any given TTL constraint, we report the configuration that maximizes system throughput, forming a unified Pareto curve.
Throughout this section, "\textbf{Baseline}" refers to the best of the baseline search space above, while "\textbf{Helix}" denotes the use of temporal pipelining with decoupled attention and FFN sharding paired with HOP-B optimization.

We constrain our search for optimal model mappings to configurations using $1$-$64$ GPUs, fitting within a single GB200 node. All model weights, KV states, and arithmetic operations are assumed to use FP4~\cite{microscaling2023}, reflecting emerging trends in low-precision LLM inference deployments.

\subsection{Results}
\begin{figure}[ht!]
  \centering
  \includegraphics[width=\linewidth]{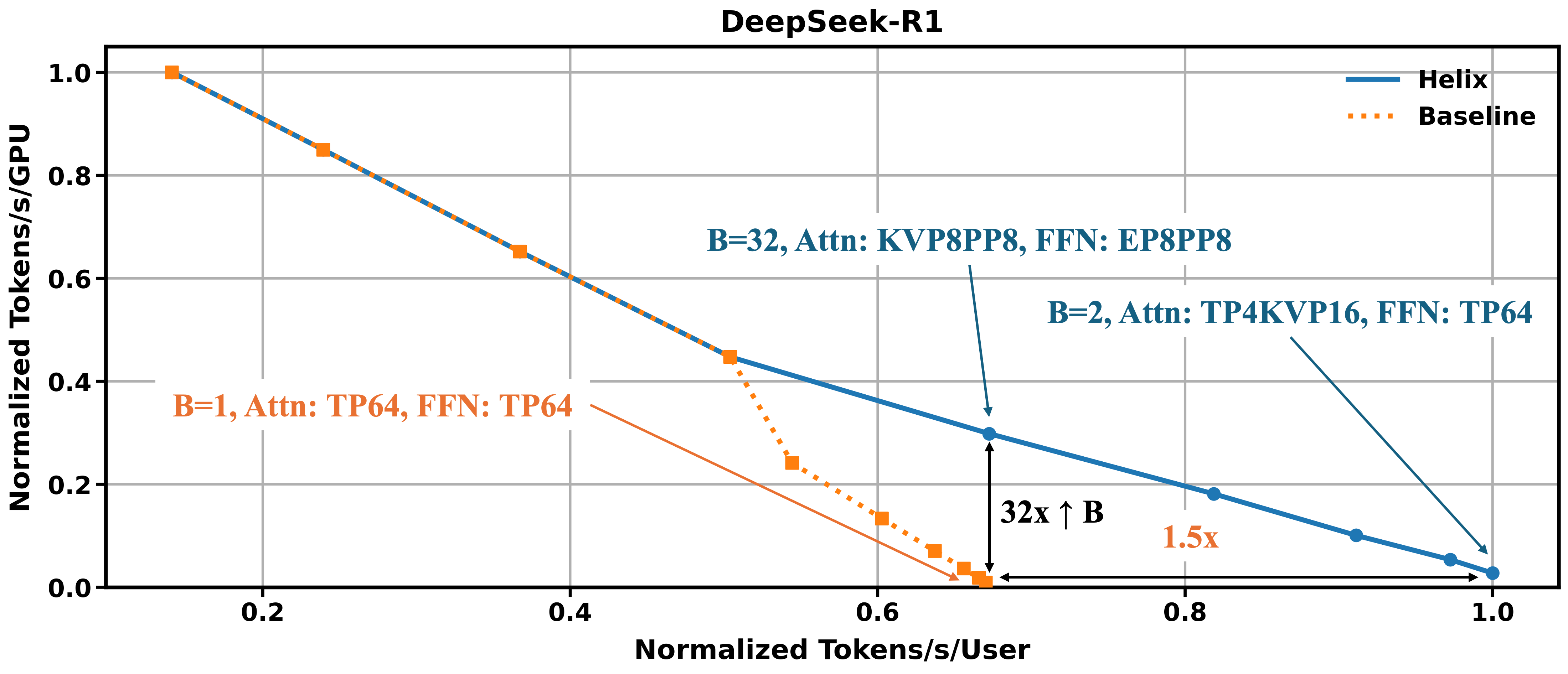}
  \caption{Pareto frontier of serving DeepSeek-R1 with 1-Million context length on GB200}.
  \label{fig:helix_pareto_deepseek}
\end{figure}

\begin{figure}[ht!]
  \centering
  \includegraphics[width=\linewidth]{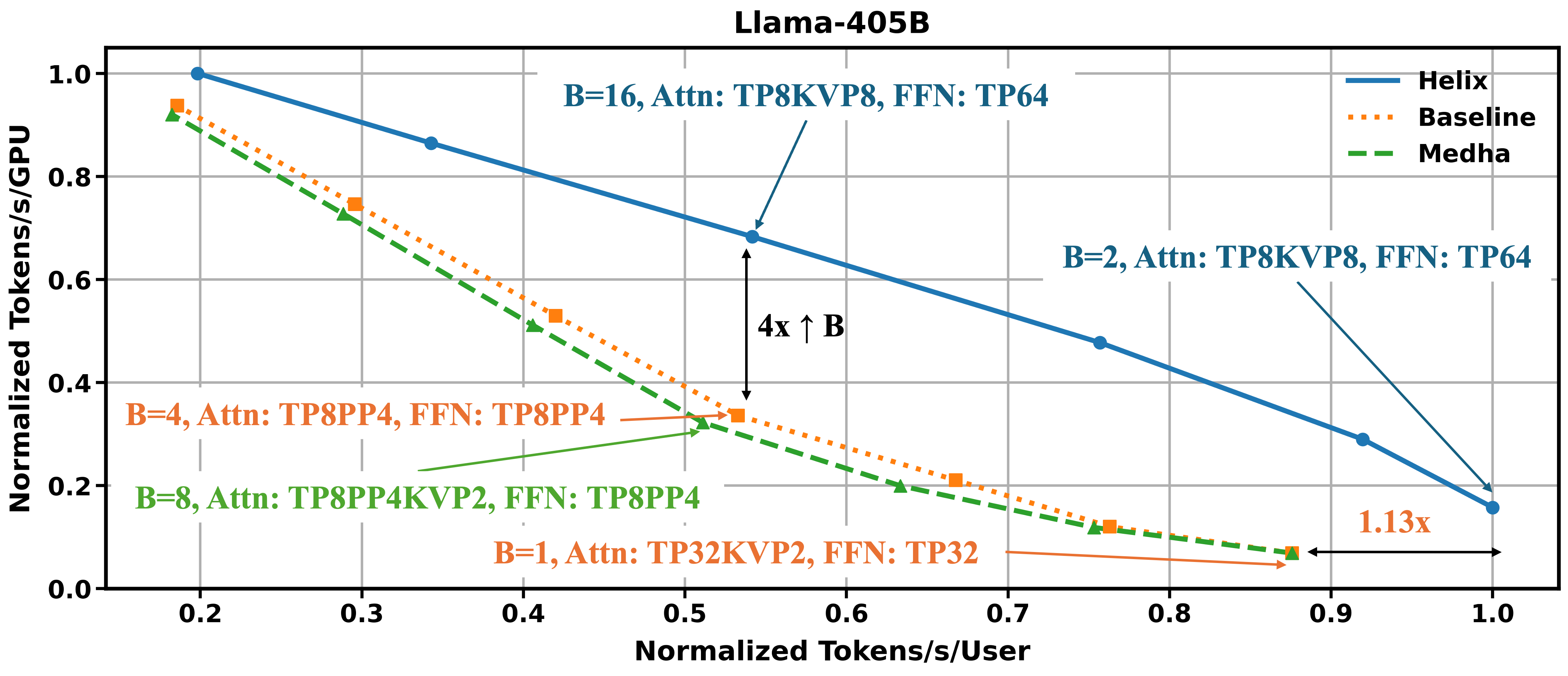}
  \caption{Pareto frontier of serving Llama-405B with 1-Million context length on GB200}.
  \label{fig:helix_pareto_llama}
\end{figure}

Figure~\ref{fig:helix_pareto_deepseek} and Figure~\ref{fig:helix_pareto_llama} provide the throughput vs. interactivity Pareto frontier for DeepSeek-R1 and Llama-405B, respectively.
To derive these Pareto frontiers, we exhaustively simulated over 100,000 configurations, systematically varying model partitioning strategies (TP, EP, PP, KVP), batch sizes, and GPU counts across different LLM architectures and inference serving techniques.\footnote{For clarity, the Pareto frontiers shown in the figures represent only the optimal configurations that achieve the best throughput-latency trade-offs.}
As demonstrated, Helix significantly improves the throughput-latency trade-off by pushing the Pareto frontier outward, enabling both higher system throughput and lower per-user latency simultaneously.

For DeepSeek-R1, Helix improves \textbf{user interactivity by up to 1.5$\times$}, enabling lower token latency while simultaneously scaling batch capacity and throughput. 
Specifically, Helix supports up to \textbf{32$\times$ more concurrent users} (i.e., achieves 32$\times$ higher Tokens/s/GPU) compared to the baseline. 
This is enabled by Helix's ability to shard both KV caches and FFN weights across all available devices, reducing DRAM pressure and increasing compute efficiency.
Note that Medha's approach of tying TP between FFNs and attention is not well-suited for modern networks with MLA attention, i.e., $TP > 1$ causes duplication of KV cache. In addition, Medha~\cite{agrawal2025medhaefficientlyservingmultimillion} does not provide any results on MoE models.
As such, a direct comparison is not applicable for the DeepSeek-R1 network.

For Llama-405B, Helix yields a \textbf{1.13$\times$ improvement in maximum achievable interactivity} and a \textbf{4$\times$ higher throughput and batch capacity} compared to TP sharding. 
These gains come from (1) lifting TP's KV-duplication ceiling via KVP, and (2) further increasing FFN parallelism without introducing cache duplication.
The comparison with Medha in Figure~\ref{fig:helix_pareto_llama} reinforces the 
benefits of removing the constraint of tying TP width between attention and FFNs. While both Helix and our baseline TP implementation include communication-computation overlap, Medha systems expose all communication overheads, which further underscores the importance of HOP-B.

\subsection{Ablation study: impact of HOP-B}
\begin{figure}[ht!]
  \centering
  \includegraphics[width=\linewidth]{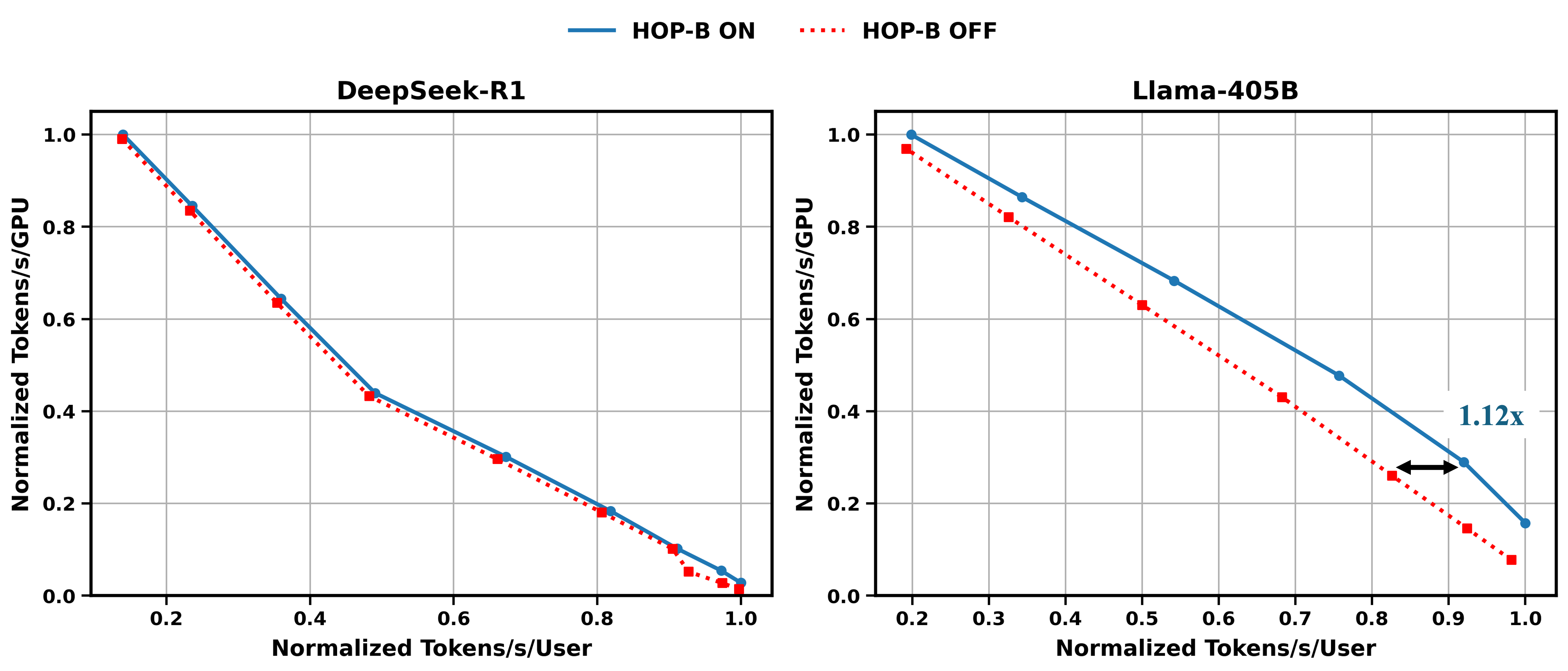}
  \caption{Pareto frontier of HOP-B ON vs. HOP-B OFF with 1-Million context length on GB200.}
  \label{fig:helix_pareto_hopb}
\end{figure}

We isolate Helix's batch-wise overlap strategy (\textbf{HOP-B}) by turning it off during attention. In "HOP-B OFF" mode, communication and computation execute strictly sequentially, incurring idle GPU stalls that reduce Tokens/s/User by up to \textbf{12\%} at a fixed Tokens/s/GPU (Figure \ref{fig:helix_pareto_hopb} \textbf{(right)}). Re-enabling HOP-B, pipelines each request's communication with the next request's attention compute, closing these gaps and recovering most of the lost TTL (Figure \ref{fig:pipeline}).

In Figure \ref{fig:helix_pareto_hopb}, DeepSeek-R1 (\textbf{left}) suffers only \textbf{\textasciitilde{}1\%} degradation when HOP-B is OFF, its all-to-all exchange accounts for just \textbf{\textasciitilde{}1\%} of end-to-end decode latency, with latent projections, shared-expert computation, and multi-expert GEMMs dominating, whereas Llama-405B (\textbf{right}) incurs a \textbf{\textasciitilde{}12\%} drop in Tokens/s/User without HOP-B. This stark contrast highlights that communication–computation overlap becomes increasingly critical as communication forms a larger fraction of TTL.

\section{Related work}
A large body of prior work has focused on sequence parallelism techniques, primarily in the context of training and prefill stages~\cite{wu2024loongserveefficientlyservinglongcontext,fang2024uspunifiedsequenceparallelism,li2024distflashattndistributedmemoryefficientattention,yang2025contextparallelismscalablemilliontoken}. 
These methods often partition the sequence dimension to improve memory efficiency and parallelism during the non-autoregressive phases of model execution. 
While effective for prefill and training, these strategies are not well-optimized for the decoding phase, where strict TTL constraints and the need for causal access 
to growing KV caches introduce unique system-level bottlenecks, particularly when dealing with KV histories spanning millions of tokens.

Another line of related work explores KV Parallelism and the practice of tying TP across attention and FFN layers to simplify execution (e.g.,~\cite{agrawal2025medhaefficientlyservingmultimillion}). 
This tight coupling, however, limits scalability especially on new model architectures such as DeepSeek-R1 which feature MLA and MoE FFNs. 

To the best of our knowledge, Helix is the first parallelism framework explicitly designed to address decoding bottlenecks in modern LLM architectures with increasing context lengths. 
By decoupling sharding strategies for attention and FFNs and introducing a temporal execution pipeline, Helix better aligns GPU utilization with the computational characteristics of each stage, enabling low latency decoding at multi-million-token context lengths.
Helix is compatible with diverse attention mechanisms, including GQA and MLA, and is co-designed with Blackwell’s latest capabilities to leverage features such as its large NVLink domain.

\section{Discussion}
Although the primarily discussion in this paper focuses on long-context inference through Helix Parallelism, it is important to note that Helix is a general-purpose inference optimization technique that extends beyond long-sequence scenarios. Its core principle, separating attention and FFN computations and distributing them independently, applies broadly across context lengths.

In the short-context regime (e.g., context lengths < 4K), Helix simplifies to data-parallel attention and tensor-parallel FFN, a pattern already widely used in modern LLM inference serving frameworks\cite{nvidiatrtllm, sglangdtp}. Helix captures a growing trend in handling inference across diverse context lengths - offering a coherent abstraction for both mainstream and emerging long-context workloads.

\section{Future work}
In this paper, we introduced and evaluated Helix parallelism and illustrated its effectiveness in optimizing decoding performance with multi-million-token KV histories across a variety of latest dense attention architectures and MoE models. One natural extension is to support sparse attention mechanisms such as Natively Sparse Attention (NSA) ~\cite{yuan2025nativesparseattentionhardwarealigned}, which reduce KV read bandwidth but not overall memory capacity requirements. Given Helix’s modular design, we expect its principles to translate naturally to these emerging architectures.

\section{Conclusion}
Helix Parallelism represents a paradigm shift in decoding efficiency for ultra-long-context LLMs operating under tight latency constraints. By decoupling parallelism as well as scaling strategies for attention and FFN layers through a temporal pipeline, Helix overcomes key bottlenecks in long-context decoding, boosting system interactivity and efficiency. Helix is fully compatible with modern LLM architectures, including diverse attention mechanisms (e.g., GQA, MLA) and sparse Mixture-of-Experts. Its design also aligns seamlessly with emerging GPU platforms such as the Blackwell system, ensuring readiness for future hardware and model trends.

\bibliographystyle{unsrt}
\bibliography{references}

\newpage
\appendix
\section{Roofline analysis}
\label{appendixA}
This section shows the formulas used to derive KV cache and FFN weight read times shown in Figure~\ref{fig:combined}.

\[B: Batch \: Size\]
\[Q: Q \: heads\]
\[K: KV \: heads\]
\[Hsz: Attention \: Head \: Size\]
\[S: KV \: Sequence \: length\]
\[H: Hidden \: dimension \: = \: Q \times Hsz\]
\[F: Intermediate \: Hidden \: dimension\]
\[TPA: TP \: width \: for \: Attention\]
\[TPF: TP \: width \: for \: FFN\]
\[KVP: KVP \: width\]
\[bytes_{param}: Bytes \: per \: parameter\]
\[MemBW: GPU \: Memory \: bandwidth\]

Time to read KV cache per LLM layer is given by below equation. It assumes individual K and V heads to represent attention variants like GQA.

\begin{equation*}
    \frac{B \times 2 \times \lceil{\frac{K}{TPA}}\rceil \times Hsz \times \frac{S}{KVP} \times bytes_{param}}{MemBW}
\end{equation*}

Time to read weights per LLM layer is given by below equation. It assumes SwiGLU activation in FFN block.

\begin{equation*}
    \frac{((2 \times H \times \frac{Q}{TPA} \times Hsz) + (2 \times H \times \lceil{\frac{K}{TPA}}\rceil \times Hsz) + (3 \times \frac{H \times F}{TPF})) \times bytes_{param}}{MemBW}
\end{equation*}

Figure~\ref{fig:combined} assumes MemBW = 8000 GB/s.

\end{document}